\documentclass[superscriptaddress,preprint,pre,aps]{revtex4}
\usepackage[utf8x]{inputenc}
\usepackage{graphicx}
\usepackage{epstopdf}
\usepackage{amsmath}
\usepackage{amsfonts}
\begin{document}
\title{Periodic ordering of clusters and stripes in a two-dimensional lattice model.
I. Ground state,  mean-field phase diagram and structure of the disordered phases}
\author{J. P\c ekalski}
\affiliation{Institute of Physical Chemistry,
 Polish Academy of Sciences, 01-224 Warszawa, Poland}
\author{ A. Ciach}
\affiliation{Institute of Physical Chemistry,
 Polish Academy of Sciences, 01-224 Warszawa, Poland}
 \author{N. G. Almarza}
\affiliation{Instituto de Qu{\'\i}mica F{\'\i}sica Rocasolano, CSIC, Serrano 119, E-28006 Madrid, Spain }
\date{\today}
\begin{abstract}
The short-range attraction and  long-range repulsion (SALR) 
between nanoparticles or macromolecules 
can lead to spontaneous pattern formation on solid 
surfaces, fluid interfaces or membranes. In order to study the  self-assembly in such systems we 
 consider a triangular lattice model with nearest-neighbor attraction and third-neighbor repulsion.
 At the ground state of the model ($T=0$) the lattice is empty  for small values of 
the chemical potential $\mu$, and fully occupied for large  
 $\mu$. For intermediate values of $\mu$
periodically distributed clusters,  bubbles or stripes appear if the repulsion is sufficiently  strong.
  At the phase coexistences  between the vacuum and the ordered cluster phases
and between the cluster and the lamellar (stripe) phases
the entropy per site does not vanish.
 As a consequence of this ground state  degeneracy,   disordered fluid phases consisting of clusters or stripes are  stable,
 and the surface tension 
vanishes. 
For $T>0$ we  construct the phase diagram in the mean-field
 approximation and calculate the correlation function in the self-consistent Brazovskii-type field theory.   

\end{abstract}
\maketitle
\section{Introduction}

Particles in many soft-matter and biological systems are charged, and repel each other with screened electrostatic 
interactions \cite{israel:11:0,shukla:08:0,stradner:04:0,campbell:05:0,seul:95:0}.
The repulsion is also present between particles covered by polymeric brushes \cite{iglesias:12:0,panagiotopoulos:13:0}
 and between membrane proteins~\cite{seul:95:0,scheve:13:0};
 in the latter case the repulsion can be caused 
by elastic deformations of the lipid membrane\cite{helfrich:73:0}. On the other hand, the particles attract
 each other with the van der Waals and 
with solvent-mediated  solvophobic, depletion or Casimir effective potentials 
\cite{shukla:08:0,iglesias:12:0,stradner:04:0,campbell:05:0,dijkstra:99:0,scheve:13:0,machta:12:0}. In addition, 
capillary forces between the particles trapped on liquid interfaces are present \cite{pergamen:12:0}.
The sum of all the interactions often has a form of the short-range attraction and  long-range repulsion 
(SALR potential)
~\cite{sear:99:0,pini:00:0,imperio:04:0,imperio:07:0,imperio:06:0,pini:06:0,archer:07:0,shukla:08:0,archer:08:0}. 

The attraction favours phase separation, while the repulsion suppresses the growth of the clusters. 
As a result, the particles can form different patterns  on surfaces, fluid interfaces or membranes. 
 The stable patterns are determined by the competition between
 the disordering effect 
of the thermal motion, the chemical potential 
controlling the number of particles, and
the attractive and repulsive parts of the interaction potential.

The  topology  of the phase diagram for particles interacting with the SALR
potential
is expected to be similar to the topology of the phase diagram in  amphiphilic systems
\cite{ciach:13:0}. The determination of the phase diagram for a particular form of the SALR potential,
however, is a real challenge both on the experimental and on the theoretical side. There are many metastable 
states and 
the time scale of ordering is large. The periods of density oscillations in different ordered phases can
be different, and  may depend on the thermodynamic state. This leads to 
incommensurability of the period of oscillations 
and the system size. The incommensurability  may strongly influence the theoretical and simulation results. 
Because of the above difficulties, the complete phase diagram for a two-dimensional (2d) system was determined
 so far  in the density-functional theory (DFT) for 
one particular shape of the SALR potential\cite{archer:08:0}. 
For the same shape of the SALR potential a sketch of the phase diagram was obtained in Ref.\cite{imperio:06:0}
 by Monte Carlo
 (MC) simulations. The potential $V(r)$ considered in Ref.\cite{archer:08:0,imperio:06:0,imperio:04:0,imperio:07:0} 
consisted of two 
 exponentially decaying terms with the  decay rates  and amplitudes (of opposite sign)  
 ensuring the global balance between the  attraction and the repulsion, i.e. $\int d{\bf r} V(r)=0$.

  In this work we are interested in the SALR potentials leading to formation of small clusters or thin stripes
 separated by distances comparable to their thickness. Such patterns can be formed when the range of
  the attraction is $\sim 1-1.5\sigma$ and the range of the repulsion is $\sim 1.5-3\sigma$, where $\sigma$
  is the particle diameter.  The above interaction ranges  are expected for cone-shape membrane
 proteins when a cluster 
of a few molecules induces a large local curvature of the lipid bilayer, and for 
  charged nanoparticls or 
  globular proteins in solvents with weak ionic strength.  In the latter systems the decay rate of the repulsion,
 i.e. the  Debye screening length, depends on the dielectric constant and the concentration of ions
 and takes the values 
   $\lambda_D\sim 1-100 nm$. Thus, the relevant particle diameters are $\sigma\sim 0.5-50 nm$. 
   The  range of  the attractive solvophobic and/or 
depletion forces   between the nanoparticles or proteins is  a little bit larger than $\sigma$.
   The above interaction ranges were found in particular for lysozyme molecules in water\cite{shukla:08:0} 
   (see Fig.1 in \cite{kowalczyk:11:0}).

  In this work we extend the lattice model introduced in Ref.
\cite{pekalski:13:0} to a 2d lattice. In order to allow for close packing of the particles, we
 consider a triangular lattice. We  postulate the interactions 
 as simple    as possible for  the above ranges of the attractive and the repulsive parts of the potential. 
We  consider various values of $\int d{\bf r} V(r)$ to study the effect of the strength of the repulsion
on the pattern formation. We pay particular 
attention to the less studied  potentials with dominant repulsion, where $\int d{\bf r} V(r)>0$. 
The calculations and simulations are much simpler in the case of
 lattice models, therefore the lattice models
 can be investigated in a great detail. Moreover, the generic lattice model can describe 
the properties common for a whole family of the SALR systems.
 We expect that for particles self-assembling  at solid substrates or on interfaces into clusters, bubbles or stripes
 the model can play analogous role   as the lattice gas (Ising) model plays 
for the phase separation.

 The model is introduced in sec.2. In sec.3 its ground state 
is described. In sec.4 the correlation function, boundary of stability of the disordered phase and the phase diagram  
are calculated in the MF approximation. In sec.5 we  describe some effects of fluctuations. 
Sec.6 contains summary and discussion.

\section{The model}
We consider a surface in equilibrium with a bulk reservoir with temperature $T$ and chemical potential $\mu_p$. The
 interaction $h$ of
 the particles with the binding sites on a solid substrate, or with the lipids in the membrane
 plays analogous role as the chemical potential, and we introduce $\mu=\mu_p+h$. 
We assume that the particles can occupy  sites of a triangular lattice with the lattice constant comparable with the 
diameter of the adsorbed particles $\sigma$. This way we allow for close packing of the particles. 
Because of this property
the triangular lattice can yield more realistic
results than the square lattice. In the case of adsorption on a solid substrate 
the model is appropriate for a triangular lattice of adsorption centers.
The lattice sites are ${\bf x}=x_1{\bf e}_1+x_2{\bf e}_2$, where  ${\bf e}_1$, ${\bf e}_2$ and
 $ {\bf e}_3={\bf e}_2-{\bf e}_1$ are the unit lattice 
vectors on the triangular lattice, i.e. $|{\bf e}_1|=|{\bf e}_2|=|{\bf e}_1-{\bf e}_2|=1$ (in $\sigma$-units),
and $x_i$ are integer. We assume
 $1\le x_i\le L$,
where $L$ is the size of the lattice in the  directions  ${\bf e}_1$ and ${\bf e}_2$. 
We also assume periodic boundary 
conditions (PBC), $L+1\equiv 1$ and $0 \equiv L$.  

\begin{figure}
 \includegraphics[scale=1]{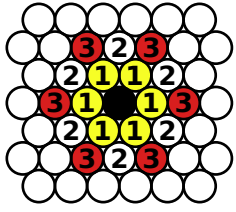}
\caption{The interactions in the lattice model. The
occupied central site (black) attracts each one of the occupied nearest-neighbor sites labelled 1 (yellow) 
and repels each one of the occupied third-neighbor sites labelled 3 (red). 
The interactions between the central site and the remaining sites vanishes. 
The strength of the attraction is $J_1$ and the strength
of the repulsion is $J_2$. 
}
\label{f1}
\end{figure}

In order to mimic the SALR interactions, we assume that the nearest-neighbors attract each other (SA), 
then the interaction changes sign for the next-nearest 
neighbors,  becomes 
repulsive for the third neighbors (LR), and vanishes for larger separations (see Fig.\ref{f1}). 
The nearest-neighbor attraction is the standard assumption in the lattice-gas models. 
In the case of charged particles in electrolyte
the assumed  range of  repulsion ($\sim 2.5\sigma$) should be  of order of the Debye screening length,
$ 2.5\sigma\sim\lambda_D$. Since
in various solvents with weak ionic strength  $\lambda_D\sim 1-100 nm$,
 the model is suitable for  charged molecules, nanoparticles or globular proteins.

The Hamiltonian has the form
\begin{equation}
\label{H}
 H= \frac{1}{2}\sum_{\bf x} \sum_{\bf x'}\hat \rho({\bf x})V({\bf x}-{\bf x}')\hat \rho({\bf x'})
-\sum_{\bf x} \mu\hat \rho({\bf x}),
\end{equation}
where $\hat\rho({\bf x})=1(0)$ when the site ${\bf x}$ is (is not) occupied. 
The interaction energy between the occupied sites  ${\bf x}$ and ${\bf x}+\Delta{\bf x}$ 
is given by
  \begin{eqnarray}
\label{V}
V(\Delta{\bf x})=\sum_{i=1}^3\Big[-J_1\Big(
\delta^{Kr}(\Delta{\bf x}+{\bf e}_i)+\delta^{Kr}(\Delta{\bf x}-{\bf e}_i)
\Big)\\
\nonumber
+J_2\Big(
\delta^{Kr}(\Delta{\bf x}+2{\bf e}_i)+\delta^{Kr}(\Delta{\bf x}-2{\bf e}_i)\Big)
\Big].
\end{eqnarray}
 $-J_1$ and $J_2$ represent the attraction well and the repulsion barrier respectively, and 
 $\delta^{Kr}({\bf x})=1$ for ${\bf x}={\bf 0}$, while $\delta^{Kr}({\bf x})=0$ for ${\bf x}\ne {\bf 0}$.

 The probability of a particular microscopic state
$\{\hat \rho({\bf x})\}$ ($\{\hat \rho({\bf x})\}$ denotes the values of $\hat \rho({\bf x})$ 
at all the lattice sites) has the  form 
\begin{equation}
\label{pB}
 p[\{\hat \rho({\bf x})\}]=\Xi^{-1}\exp(-\beta H[\{\hat \rho({\bf x})\}]),
\end{equation}
where $\beta=1/(k_BT)$ and $k_B$ is the Boltzmann constant. The grand potential is 
expressed in terms of the grand statistical sum 
\begin{equation}
\label{Xi}
 \Xi=\sum_{\{\hat \rho({\bf x})\}}\exp(-\beta H[\{\hat \rho({\bf x})\}])
\end{equation}
in the standard way
\begin{equation}
\label{Omdef}
 \Omega=-k_BT\ln \Xi=-pa_{0}L^2=\langle H\rangle -TS=U -TS-\mu\langle N\rangle,
\end{equation}
where $a_0=\sigma^2\sqrt 3/2$ is the area per lattice site, 
 $p$ is 2d pressure, $\langle N\rangle$ is the average number of particles, $S$ is the entropy, and  
 the internal energy 
 is  $U=\langle  H +\mu N \rangle$. 

 The  probability of the state $\{\hat \rho({\bf x})\}$ for $\mu=\tilde V(0)/2-\Delta\mu$ is the same as 
the probability of the  state $\{1-\hat \rho({\bf x})\}$ for $\mu=\tilde V(0)/2+\Delta\mu$ \cite{pekalski:13:0}, where
\begin{equation}
\tilde V(0)=\sum_{\bf x} V({\bf x})=6(J_2-J_1).
\end{equation}
Because of the above property, the phase diagram is symmetric with respect to the symmetry axis 
$\mu=\tilde V(0)/2=3(J_2-J_1)$.

We choose $J_1$ as the energy unit, and introduce the notation  $X^*=X/J_1$  for any quantity $X$ with 
dimension of energy, in particular
\begin{equation}
\label{dimensionless}
T^*=k_BT/J_1,\quad J^*=J_2/J_1, \quad \mu^*=\mu/J_1, \quad H^*=H/J_1.
 \end{equation}

\section{The ground state}
\begin{figure}[ht]
\includegraphics[scale=1]{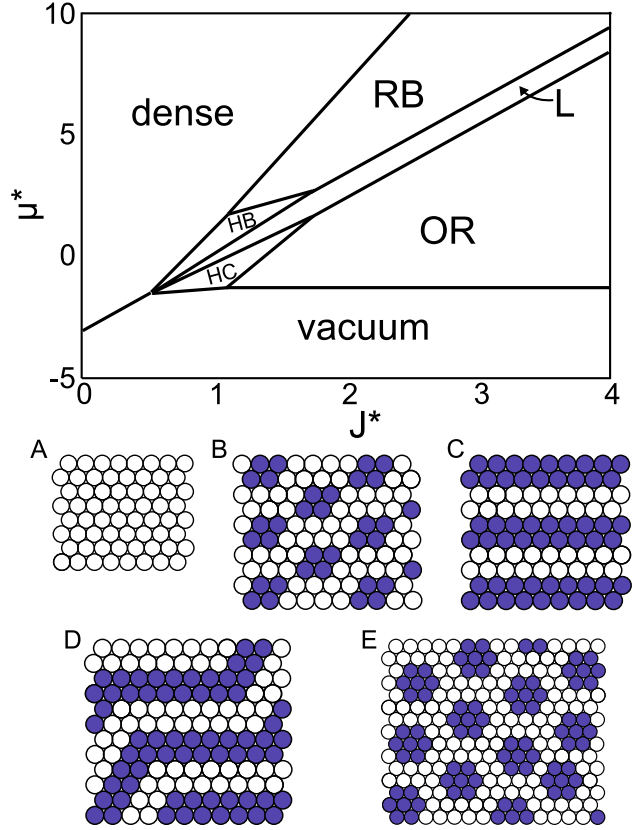}
\caption{Ground state of the model. $\mu^*$ and $J^*$ denote the chemical potential and the third-neighbor repulsion
respectively, both in units of the nearest-neighbor attraction. The structures of the stable phases are shown 
in the panels A-E, with: A) vacuum, B) ordered rhomboidal clusters (OR), C) and  D) lamellar phase (L), E) 
hexagonal clusters (HC).  Dense phase,  hexagonal bubble phase (HB) and rhomboidal bubble phase (RB) 
are ''negatives`` of the phases a), b) and c) respectively, i.e. the occupied sites are replaced by 
the empty ones and vice versa. The symmetry line is given by $\mu^* = 3J^*-3$. 
Configurations stable at the coexistence lines are shown in Fig. \ref{gsc}.}
\label{gs}
\end{figure}

\begin{figure}[ht]
\includegraphics[scale=1]{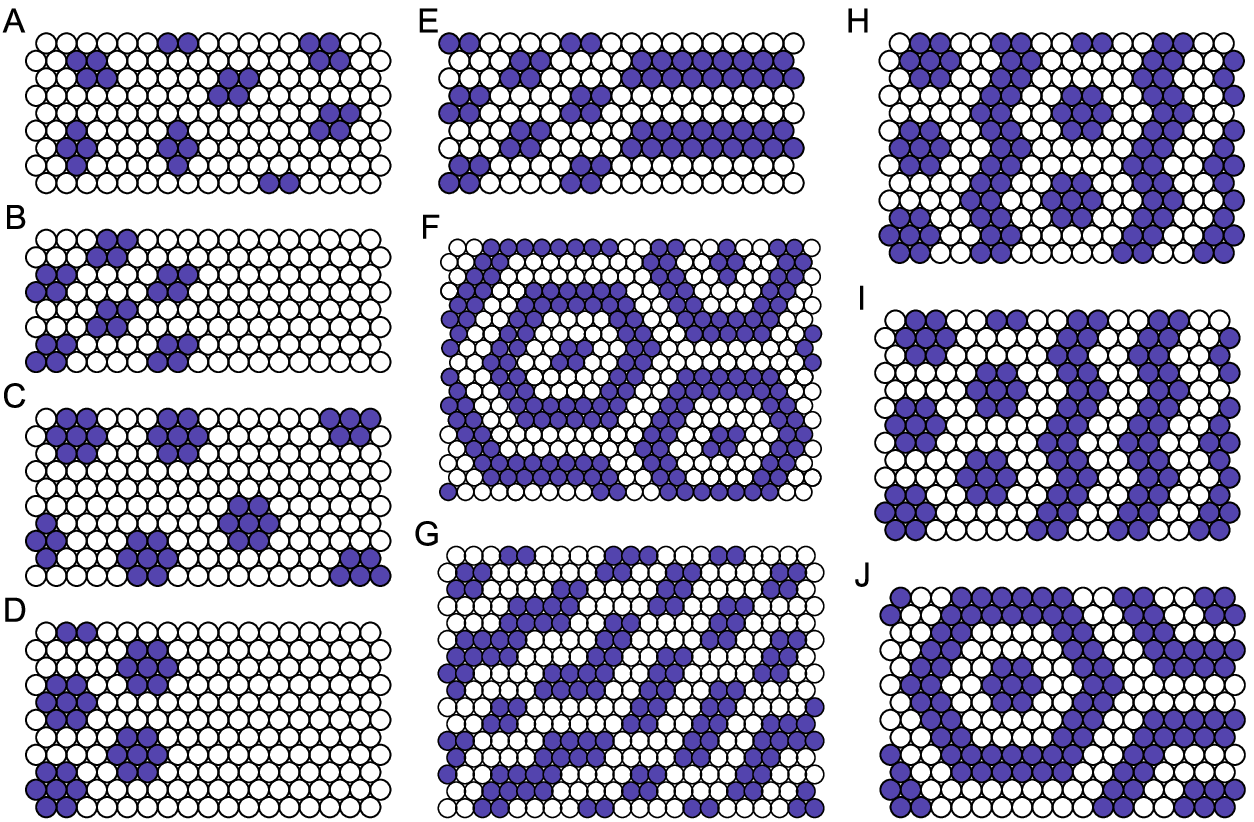}
\caption{Examples of the ground state structures stable at the coexistence lines. Left panels: vacuum - OR phase 
coexistence (a-b) and vacuum - HC phase coexistence (c-d).
Central panels: OR - lamellar phase coexistence.
Right panels: HC - lamellar phase coexistence. }
\label{gsc}
\end{figure}

The grand potential  for $T=0$ reduces to the minimum of $H^* [\{\hat\rho(x)\}]/L^2$.
The stability regions of the homogeneous
and various periodic phases on the $(J^*,\mu^*)$ plane were obtained by a direct calculation of
 $H^* [\{\hat\rho(x)\}]/L^2$. 
Two phases can coexist when $H^* [\{\hat\rho(x)\}]/L^2=-p^*$ in these phases takes the same value. 
The ground state (GS) and the structure of the stable phases are shown in Fig.\ref{gs} and \ref{gsc}.
For weak repulsion the vacuum and the fully occupied 
lattice coexist for $\mu^*=3J^*-3$. For $J^*>1/2$ the stability regions of the two phases are separated by the 
region of stability of periodic structures. The  topology of the ground state is similar to the one found before
in the 1d version of the model
\cite{pekalski:13:0}, except that the stability region of the periodic phase splits into stability regions of several 
periodic phases: hexagonally ordered clusters of rhomboidal (OR) or hexagonal (HC) shape, the stripe (lamellar) phase
 (L) and hexagonally ordered rhomboidal (RB) or hexagonal (HB) bubbles. By the model symmetry, the bubble phases
 are ``negatives`` (i.e. $\hat\rho({\bf x})\to 1-\hat\rho({\bf x})$) of the cluster phases.

For $J^*>1/2$ the ground state is strongly degenerated at the coexistence lines,
and the entropy per lattice site does not vanish.
This can be easily shown for the coexistence between the vacuum and the OR or HC phases.  
In the vacuum $H^*=0$.
The change of $H^* [\{\hat\rho({\bf x})\}]$ when a single rhomboidal or   hexagonal cluster appears in the vacuum is
 $-5-4\mu^*$ or $-12+3J^*-7\mu^*$, respectively.
 For $\mu^*=-5/4$ or $\mu^*=(-12+3J^*)/7$ the Hamiltonian does not change if an arbitrary number of 
noninteracting rhomboidal or hexagonal clusters  appears in the system. 
The  clusters do not interact   if the distance between them is sufficiently large. 
There are no more
restrictions on the  positions and orientations of the clusters in the states with
  $H^*=0$  at the vacuum - OR and vacuum - HC phase coexistences (Fig.\ref{gsc}a,c). 
For this reason the entropy per lattice site does not vanish.
 Note that the surface tension between the vacuum and the OR phases as well as
 between the vacuum and the HC phases vanishes,
because  $H^*=0$ when the interface between the two phases is present (Fig.\ref{gsc}b,d). For $-5-4\mu^*<0$ and
 $-12+3J^*-7\mu^*<0$ the OR and HC phases respectively
 are more stable than the vacuum. In these phases the noninteracting clusters are
 packed as densely as possible (Fig.\ref{gs} b and e).

 The HC and OR phases coexist with the lamellar phase for $\mu^*=(13J^*-14)/5$ and $\mu^*=3J^*-7/2$ respectively. 
 Note that $p^*$  takes the same value in the lamellar phase shown in Fig.\ref{gs}c,
 and in the zig-zag lamellar phase shown in Fig.\ref{gs}d. 
There are many configurations 
of the zig-zag  stripes having thickness 2  in one of the lattice directions and separated by empty regions of the 
same shape (Fig.\ref{gs} d).
Thus, in the stability region of the lamellar phase the
 GS is degenerated. The zig-zag lamellas are discussed in more detail in Ref.\cite{almarza:14:0}.  
 
 At the coexistence between the lamellar and the ordered cluster phases there exists
a large number of disordered states with the same value of $p^*$ as for the two coexisting ordered phases.
Characteristic examples of such states are shown in Fig.\ref{gsc} e-j. 
 Note that these states include the interface between the ordered cluster and the lamellar phases (Fig.\ref{gsc} e,i).
In Fig.\ref{gsc}g closely packed zig-zag clusters of
different length are present. The thickness of the clusters in  direction ${\bf e}_3$
 is 2 except at the two opposing vertices where the thickness is 1. In Fig.\ref{gsc}f,j the clusters are surrounded by
lamellar rings. Structures with a few closely packed clusters surrounded by one or a few lamellar rings are stable too.
All the clusters, layers or rings
 are packed as densely as possible under the constraint that  
 the neighboring objects do not repel each other. More precisely, 
  the polygons obtained by  surrounding the clusters or stripes by  a single 
layer of empty sites must cover the whole lattice. 
This requirement follows from the negative value of the grand potential per site in the L, HC and OR phases.
We call the phase stable along the coexistence between the lamellar and the HC or OR phases  a 'molten lamella'.
The GS at the HC - OR coexistence, $\mu^*=(36J^*-49)/8$, is not degenerated.

The degeneracy of the GS at the phase coexistence 
 and the vanishing surface tension are closely related. An arbitrary number 
 of interfaces can appear when the surface tension  vanishes. As a result, 
the number and the size of the droplets of the 
 coexisting phases can be 
 arbitrary. This leads to disordered states that can be considered as fluids of clusters or stripes.
 At $T^*=0$ these disordered phases  are stable only at the phase coexistence, i.e.
 for a single value of 
 the chemical potential for given
 interaction strength.

\section{MF approximation}

We consider stable or metastable structures with densities periodic in space. 
For the position-dependent density $\bar \rho({\bf x})$ the  mean-field acting on the 
site ${\bf x}$ has the form
\begin{equation}
 \label{h}
h({\bf x})=-\sum_{{\bf x'}}V({\bf x}-{\bf x'})\bar\rho({\bf x'}),
\end{equation}
where the interaction potential $V$ is defined in Eq.(\ref{V}).
The MF grand potential is
\begin{eqnarray}
\label{wzomeg}
 \Omega_{MF} &=&\frac{1}{2}\sum_{{\bf x}_1}\sum_{{\bf x}_2}\bar\rho(
{\bf x}_1)\bar\rho({\bf x}_2) V({\bf x}_1-{\bf x}_2) + \sum_{\bf x} f_h(\bar \rho({\bf x}))
-\mu\sum_{\bf x}\bar\rho({\bf x}),
\end{eqnarray}
where
\begin{equation}
\label{fh}
f_h(\bar \rho({\bf x})) = k_{B}T \Big[ \bar\rho({\bf x})\ln (\bar \rho({\bf x}))+(1- \bar\rho({\bf x}))\ln(1-
 \bar\rho({\bf x}))
\Big].
\end{equation}
The grand potential (\ref{wzomeg})
 assumes a  minimum for $\bar\rho$ which satisfies the self-consistent equation \cite{ciach:11:1,pekalski:13:0}
\begin{equation}
\label{ror}
 \bar\rho({\bf x})=\frac{e^{\beta(h(
{\bf x})+\mu)}}{1+e^{\beta(h({\bf x})+\mu)}}.
\end{equation}
\subsection{The structure of the disordered phase} 
The structure factor  in the disordered phase (with $\bar\rho=const.$) 
 is obtained from the relations  $S({\bf k})=\tilde G({\bf k})/\bar\rho $ and
 $\tilde G({\bf k})=1/\tilde C({\bf k})$ \cite{evans:79:0}. 
In MF $\tilde G_{MF}({\bf k})=1/\tilde C_{MF}({\bf k})$, where 
\begin{equation}
\label{Gk}
 \tilde C_{MF}({\bf k})=\frac{\delta^2 \beta\Omega_{MF}}{\delta \tilde\rho({\bf k})\delta \tilde\rho(-{\bf k})}
 =\beta\tilde V({\bf k})+\frac{1}{\bar\rho(1-\bar\rho)}.
\end{equation}
In the above $\tilde\rho({\bf k})=\sum_{\bf x} \rho({\bf x})e^{i{\bf k}\cdot {\bf x}}$ and
\begin{eqnarray}
\label{Vk}
 \beta \tilde V({\bf k})
=\sum_{\bf x} \beta V({\bf x})e^{i{\bf k}\cdot {\bf x}} =
\nonumber
\\
2\beta^*\Big[J^*\Big(\cos (2k_1)+\cos (2k_2)+\cos(2(k_1-k_2))\Big)-
\cos k_1-\cos k_2-\cos(k_1-k_2)
\Big]
\end{eqnarray}
is the interaction potential in the Fourier representation. In the case of the triangular lattice
${\bf k}\equiv(k_1,k_2)=k_1{\bf f}_1+k_2{\bf f}_2$, and ${\bf x}\cdot{\bf y}$ is the standard scalar product
in $\mathbb{R}^2$. 
The unit vectors of the dual lattice satisfy $ {\bf f}_i\cdot{\bf e}_j=\delta^{Kr}_{ij}$  and $ |{\bf f}_i|=2/\sqrt 3$. 

\begin{figure}
\includegraphics[scale=1]{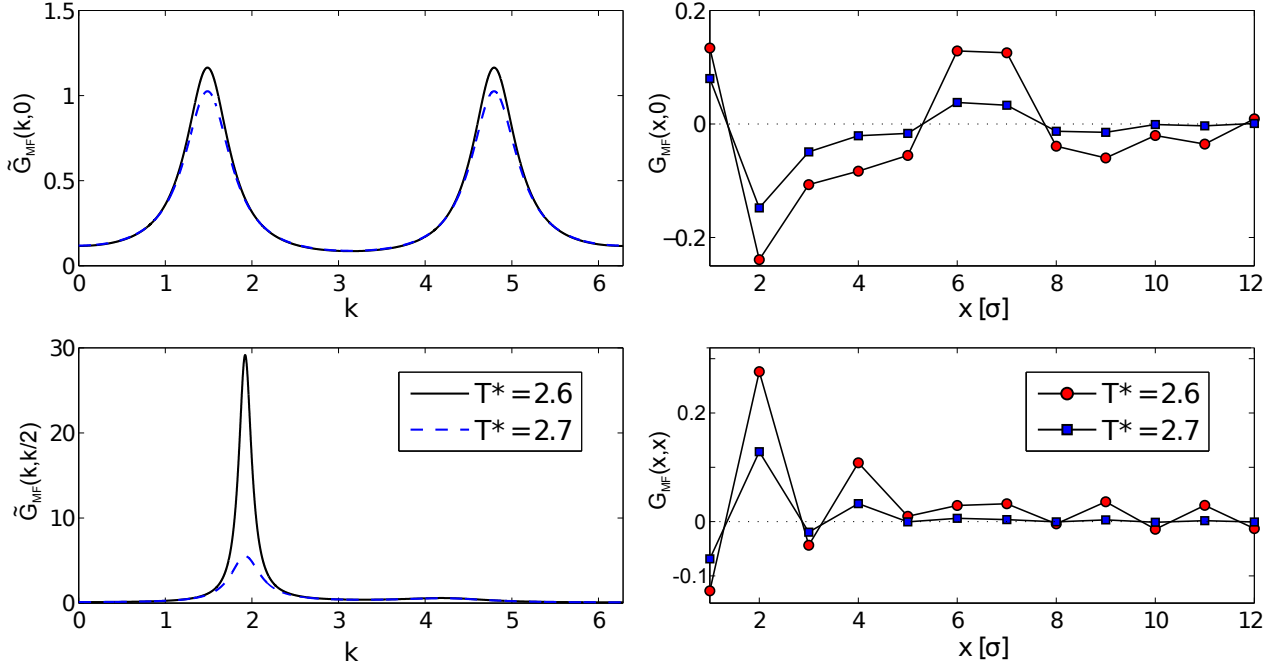}
\caption{The correlation function for $J^* = 3$ and $\rho^* = 0.5$. Red circles and solid lines correspond to 
$T^* = 2.6$ while blue squares and dashed lines to $T^* = 2.8$. 
Left column: G in Fourier space;  top panel: $\tilde G_{MF}(k,0)$, bottom panel:
$\tilde G_{MF}(k,k/2)$. Right column: G in real space; top panel:  $ G_{MF}(x,0)$, i.e. for points ${\bf x}=x{\bf e}_1$,
 and bottom panel:  $G_{MF}(x,x)$, i.e. for points ${\bf x}=x{\bf e}_1+x{\bf e}_2$. 
 The temperature of the $\lambda$- line is $T^*_{\lambda} = 2.575$.  }
 \label{G3}
\end{figure}

\begin{figure}
\includegraphics[scale=1]{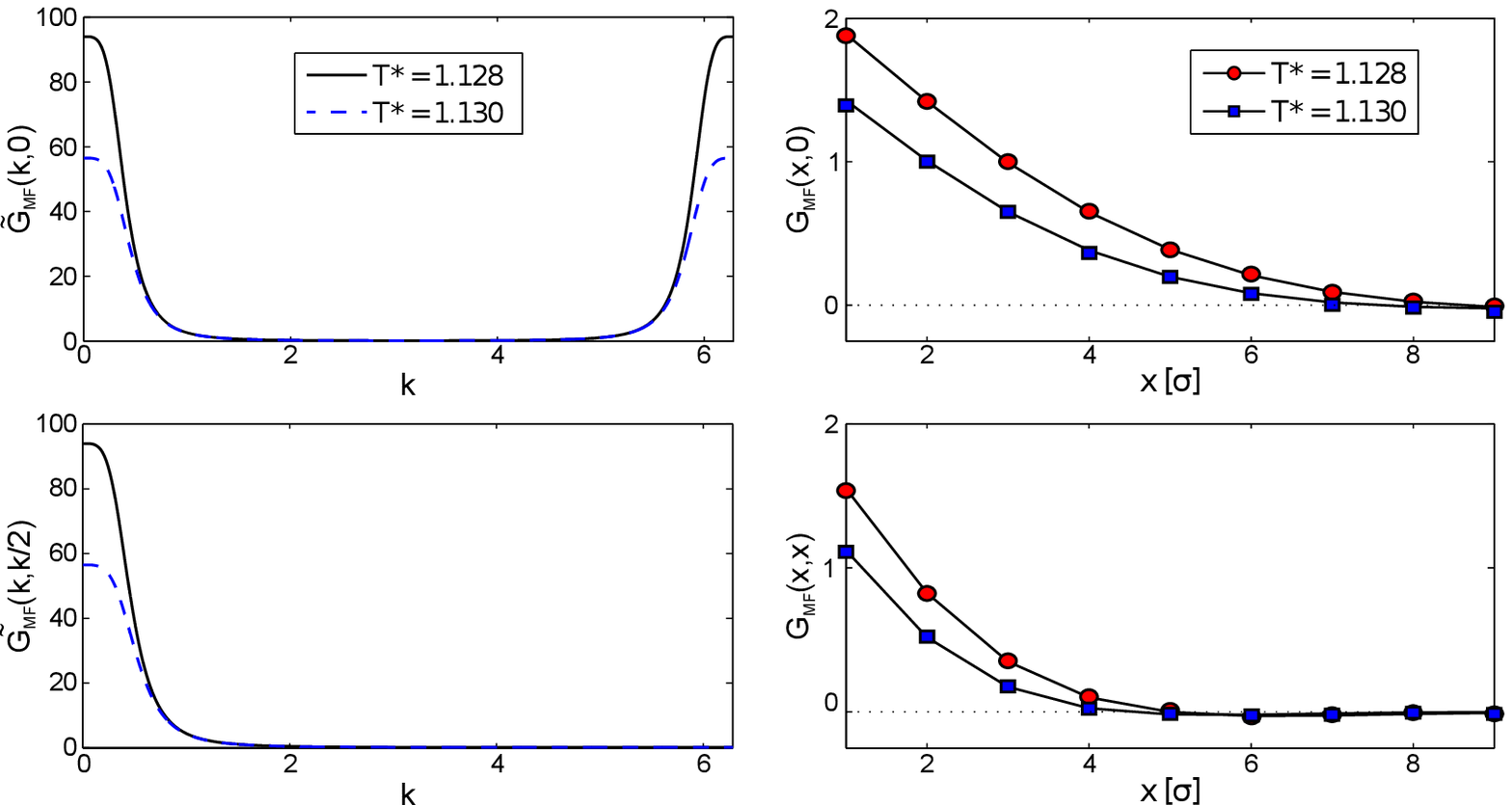}
\caption{The correlation function for $J^* = 1/4$ and  $\rho^* = 0.5$. 
  Red circles and black solid lines correspond to 
$T^* = 1.128$ while blue squares and dashed lines to $T^* = 1.13$. 
Left column: G in Fourier space;  top panel: $\tilde G_{MF}(k,0)$, bottom panel:
$\tilde G_{MF}(k,k/2)$. Right column: G in real space; top panel:  $ G_{MF}(x,0)$, i.e. for points ${\bf x}=x{\bf e}_1$,
 and bottom panel:  $G_{MF}(x,x)$, i.e. for points ${\bf x}=x{\bf e}_1+x{\bf e}_2$. 
 The temperature of the $\lambda$- line is $T^*_{\lambda} = 1.124$. }
 \label{G1/4}
\end{figure}

The maximum of the structure factor $S({\bf k})$ corresponds to the minimum of $\tilde V({\bf k})$.
 For $J^*< 1/4$
the function given by Eq.(\ref{Vk})  assumes the 
minimum for ${\bf k}={\bf 0}$, whereas for $J^*\ge 1/4$ the minimum occurs for
$k_1=2 k_2=k_b$ with 
\begin{equation}
\label{kb}
 k_b=2 \arccos \Big(\frac{J^*+\sqrt{J^{*2}+2J^*}}{4J^*}\Big).
\end{equation}
(In Ref.\cite{ciach:11:1} this extremum of
$\tilde V({\bf k})$ was overlooked.) By symmetry of the lattice there are two other minima of the same depth.
Thus,  $\tilde V({\bf k})$ takes the global minima for the wavevectors 
\begin{equation}
\label{KB}
 {\bf k}_b^{(i)}=k_b{\bf e}_i.
\end{equation}
 We used the relations $ {\bf e}_1={\bf f}_1+\frac{1}{2}{\bf f}_2$ and $ {\bf e}_2={\bf f}_2+\frac{1}{2}{\bf f}_1$.
Note that the  characterisitc length $2\pi/k_b$ is noninteger. Thus, the period of damped oscillations
 in the correlation
function is incommensurate with the lattice. Similar result was obtained by the exact transfer matrix method
 for the 1d version of our model \cite{pekalski:13:0}.  
 In Figs.\ref{G3} and \ref{G1/4} we show the correlation function $G_{MF}$ in Fourier and real-space representation
 for $J^*=3$ and $J^*=1/4$ respectively. 
 
 \subsection{Boundary of stability of the disordered phase}
The disordered  phase is unstable 
 if the grand potential decreases when the density wave with an infinitesimal amplitude
 and some wavevector
 ${\bf k}$ appears, i.e. when $\tilde C_{MF}({\bf k})<0$.  The  boundary of stability of the disordered phase 
is given by 
 $\tilde C_{MF}({\bf k}_b)=0$. For $k_b=0$ and $k_b>0$ it corresponds to the spinodal and the $\lambda$-line 
respectively. From Eq.(\ref{Gk})
we obtain the explicit expression for the boundary of stability of the disordered phase
\begin{equation}
\label{Tlambda}
T^*_{\lambda}=-\tilde V^*({\bf k}_b)\rho(1-\rho).
\end{equation}
 In the density waves that destabilize the disordered phase the density oscillates in the principal
  directions of the lattice (see (\ref{KB})).
In the case of the lamellar structure  the layers of constant density are perpendicular to either one of the
 unit lattice vectors
 ${\bf e}_i$ (three-fold degeneracy). In the case of the hexagonal structure the density is a superposition of 3 
 planar density waves in the principal lattice directions.
 
\begin{figure}
\includegraphics[scale=1]{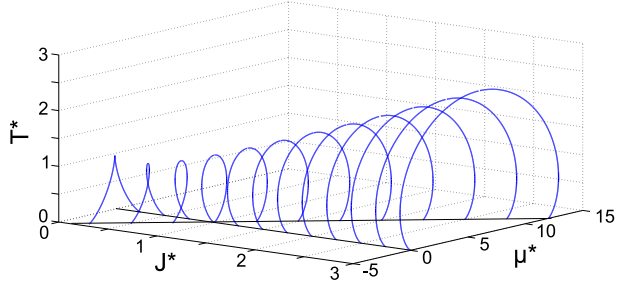}
\caption{MF lines of instability (solid) of the homogeneous phase in the $(\mu^*,T^*)$ variables for a range of $J^*$.
 We used $\partial \Omega_{MF}/\partial \rho=0$, (\ref{wzomeg})  and (\ref{Tlambda}). }
\label{fig_spin}
\end{figure}

 The shape of the $\lambda$-line in the $(\rho,T^*)$ variables is the same as the shape of the spinodal
 line of the phase separation, except that 
 the temperature scale is given by $-\tilde V({\bf k}_b)$ rather than by $-\tilde V({\bf 0})$. 
 This property is common for different forms of the SALR potential \cite{archer:08:0,ciach:08:1,pekalski:13:0}. 
 However, in $(\mu^*,T^*)$ variables the shapes 
 of the spinodal and the $\lambda$-lines differ significantly from each other. Moreover,
 the shape of the $\lambda$-line
 depends on $J^*$  (Fig.\ref{fig_spin}). For $J^*<1/4$ the  two branches of the spinodal form a cusp. 
 On the low-$T^*$ side of these lines there are two minima of $\Omega_{MF}^*$, corresponding to the
 gas and liquid phases.
 For $J^*>1/4$  the two branches form a loop for high $T^*$. 
  Inside the loop the grand potential assumes a minimum for periodic structures. 
  For increasing $J^*$ the size of the loop increases,
 and for $J^*>1$ the gas- and liquid branches of the instability line disappear.
 Similar shapes were obtained
 in the one-dimensional lattice model \cite{pekalski:13:0} and in the three-dimensional continuous 
model \cite{barbosa:93:0}.
 Thus, the above evolution of the MF lines of instability for increasing repulsion seems to be a generic property, 
 independent of the particular shape of the SALR potential and dimensionality of the system.  Note that for $J^*>1/4$ 
we obtain instability with respect to periodic ordering for high $T^*$ in MF, whereas for
 $T^*=0$ the periodic phases appear only for $J^*>1/2$. Thus, for $1/4<J^*<1/2$ gas and liquid phases are stable
 for low $T^*$,
periodic structures occur for intermediate $T^*$, and for high $T^*$ a disordered phase is stable. 
Phase separation for low $T^*$ and periodic 
ordering for high $T^*$ was
observed before for different forms of the SALR potential for moderate repulsion in MF theories
\cite{pekalski:13:0,andelman:87:0,archer:08:1}.
 
 \subsection{First-order transitions}

We solve Eq.(\ref{ror}) numerically by iterations with initial states of different symmetries and periods,
 and next compare the MF grand potential (\ref{wzomeg})  per lattice site for the obtained metastable structures.
 We assume PBC and consider different values of $L$. This way structures with periods $L/n$ where $n$ is integer 
can be generated.  
We find very large number of metastable states, especially for high $T^*$, where the order is weak 
(small amplitude of the density oscillations).
To overcome this problem we assume that when  the amplitudes of the density
oscillations in the
  periodic phases   are small,  the density has the form
\begin{equation}
\label{rhop}
 \rho_p({\bf x})=\rho_0+\delta\rho_p+\Phi_pg_p({\bf x}).
\end{equation}
In the above  
$\rho_0$ is the  position-independent density corresponding to  the extremum of $\Omega_{MF}$ 
for given $\mu^*$ and $T^*$. The $\delta\rho_p$
 is the shift of the average density in the periodic phase $p$, and $g_p({\bf x})$
is the normalized periodic function with the symmetry of the corresponding $p$ phase, 
where $p=l,h$   for the lamellar and the hexagonal phase respectively.
For the densities of the form (\ref{rhop}) the excess grand potential,
\begin{equation}
\label{delomp}
 \Delta\Omega_p[\rho_p]=\Omega_{MF}[\rho_p]-\Omega_{MF}[\rho_0],
\end{equation}
 is a function of $\delta\rho_p$ and $\Phi_p$ (see Eq.(\ref{wzomeg})). 
It  takes a minimum for  $\delta\rho_p$ and $\Phi_p$ corresponding to a stable or a metastable phase $p$.
We limit ourselves to $\delta\rho_p\to 0$ and $\Phi_p\to 0$, and from the conditions 
$\partial \Delta\Omega_p/\partial \delta\rho_p=0=\partial \Delta\Omega_p/\partial\Phi_p$ obtain the approximate values 
of $\delta\rho_p$ and $\Phi_p$, and of the excess grand potential in the lamellar and hexagonal phases.
Next, from $ \Delta\Omega_h=0$ and $ \Delta\Omega_h=\Delta\Omega_l$ we obtain the transitions between
 the disordered and hexagonal, and  between the hexagonal and the lamellar  phases respectively. 
These transition lines are shown as dashed lines in Fig.\ref{MFpd}. 
Some details of the calculation are given in Appendix.

The phase diagram obtained in the MF approximation described above  is presented in Fig. \ref{MFpd} for $J^*=3$.
 F, H, OR, L$_1$ and L$_2$ denote the disordered fluid, the high-$T^*$ hexagonal phase, 
the ordered rhombus, 
and the low-$T^*$ and high-$T^*$ lamellar phases respectively. 
The MF density distribution in the H and L$_2$ phases is shown in 
Fig. \ref{cf2}, and the structure of the 
OR and  L$_1$ phases for $T^*\to 0$ is shown in Fig.\ref{gs}b 
and Fig.\ref{gs}c,d respectively. 
In the H phase the clusters form a hexagonal pattern, but in contrast to the OR phase the orientation of the 
long axes of the rhombuses is not fixed. 
The OR phase can be present only in the case of small asymmetric clusters, i.e. for large repulsion.
 For $J^*=1$ (hence for $\int d{\bf r}V(r)=0$) hexagonal clusters
appear for $T^*=0$ (Fig. \ref{gs} e) and only positional ordering of the clusters can occur. 
In the L$_2$ phase the orientation of the
 lamellas differs from the ground-state
 orientations, and agrees with the orientation of the density waves that destabilize the homogeneous phase. 
Beacuse of a very large number of metastable structures characterized by very similar values of the grand potential,
it is likely that some details of the  phase diagram 
are not reproduced in Fig.\ref{MFpd}
 with full precision.

\begin{figure}
\includegraphics[scale=1]{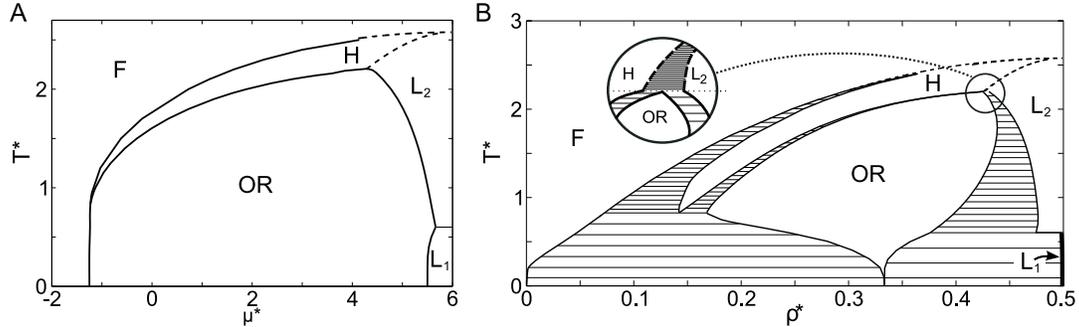}
\caption{Mean-field phase diagram for $J^* = 3$ in ($\mu^*, T^*$) (Panel A) and ($\rho^*,T^*$)
 (Panel B) variables. F, H, OR, L$_1$ and L$_2$ denote disordered fluid,
 hexagonal phase, ordered rhombuses and low- and high temperature lamellar 
phases respectively. Typical microstates of the 
  OR and  L$_1$ phases are shown in Figs.   \ref{gs}b 
 and  \ref{gs}c,d  respectively. MF density profiles in the phases H and L$_2$ are
 shown in Fig.\ref{cf2}. 
 The L$_1$ lamellar phase  is stable for $T^* < 0.65$ and $\rho \approx 0.5$ (the density interval 
is  within the thickness of the line). When temperature rises ($T^*> 0.65$), the L$_2$ phase (see Fig.\ref{cf2}b)
  becomes stable. 
The density ranges of the two-phase regions for temperatures $T^* > 2$ (dashed lines) are also within
 the thickness of the line.
 }
\label{MFpd}
\end{figure}
\begin{figure}
\includegraphics[scale=1]{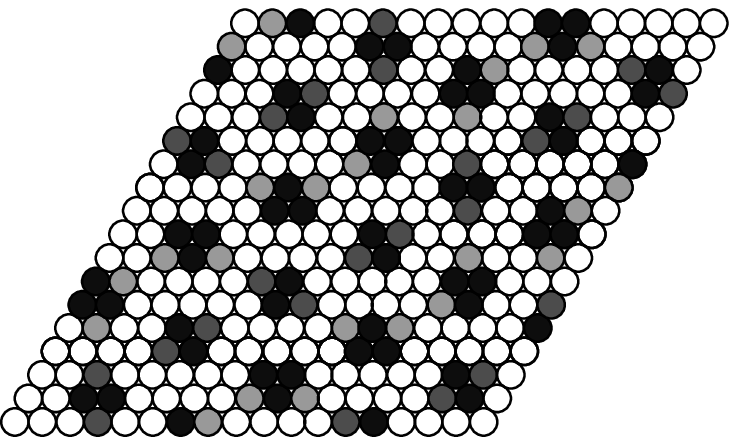}
\includegraphics[scale=0.9]{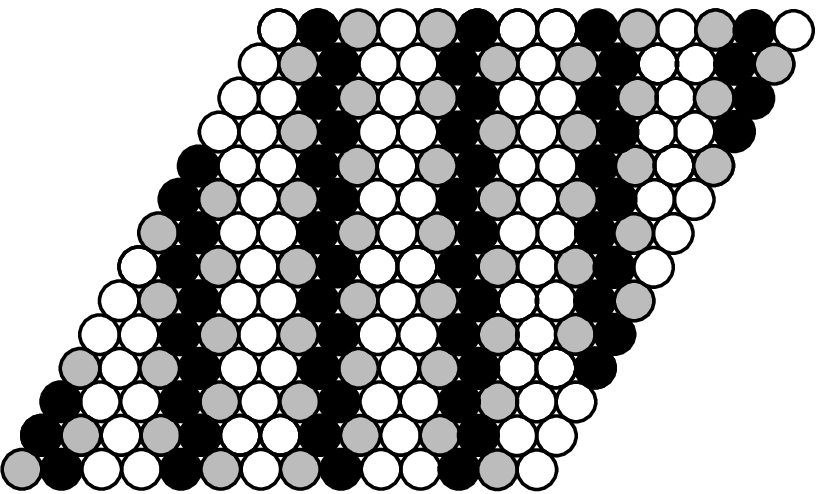}
\caption{A: Structure of the hexagonal phase H for $J^*=3$, $T^* = 0.925$ and $\rho^* = 0.15364$. 
Site colors represent different 
ranges of probability for a particle to occur at the given site, namely: white - $(0,0.065)$, 
light gray - $(0.206,0.213)$ , dark gray - $(0.37,0.4)$, black - $(0.5,0.57)$. 
B: Structure of the lamellar phase L$_2$  for $J^*=3$,  $T^* = 1$ and $\rho^* = 0.4905$.
 Sites colors represent different 
ranges of probability for a particle to occur at the given site, namely: white - $(0,0.045)$, 
light gray - $(0.6758,0.676)$, black - $0.9955$. }
\label{cf2}
\end{figure}

\section{Beyond Mean Field}

In MF the self-assembled clusters and stripes  are  present only in the ordered  phases,
 and  for $T \to 0$ the density of the disordered phase at the coexistence with the OR phase  is very small.
 This is because in the case of delocalized clusters the average density  is position independent, and 
the repulsion contribution to the mean-field grand potential
for large 
position-independent 
density is  large (see (\ref{wzomeg})). In the case of rhomboidal clusters separated by  distances larger 
than the range of the repulsion, however,
the repulsion contribution to the internal energy is absent.
Therefore for low $T^*$ the density of the disordered phase at 
the coexistence with the ordered 
cluster phase is significantly underestimated in the mean-field approximation.
In  sec.5a we take into account the degeneracy of the GS and present a semi-quantitative analysis of 
the disordered  cluster fluid  for $\beta^*(5+4\mu^*)\to 0$ and $\beta^*\gg 0$.

For  high $T^*$ thermal fluctuations destroy the periodic order, and the stability region of the disordered phase
 enlarges compared to the MF results. Based on the results obtained 
earlier for similar models \cite{pekalski:13:0,archer:08:1,imperio:06:0}  
we expect that the phases with small amplitude of the density oscillations posses only  short-range order beyond MF.
We thus expect locally hexagonal arrangement of clusters instead of the  H  phase
and locally lamellar order  instead of the L$_2$  phase for $T^*>1.5$.
According to the Brazovskii theory \cite{brazovskii:75:0}, the order-disorder transition 
to a lamellar phase is fluctuation-induced 
first order in off-lattice systems. On various lattices, however, either first order or continuous order-disorder 
transition to the periodically ordered phase may occur \cite{ciach:03:0}. 
We determine the order of this transition,
 and calculate the correlation function in the field-theoretic formalism in sec. 5b.
 
\subsection{The effect of the degeneracy of the ground state on the phase diagram for $T^*>0$}

\begin{figure}
\includegraphics[scale=0.8]{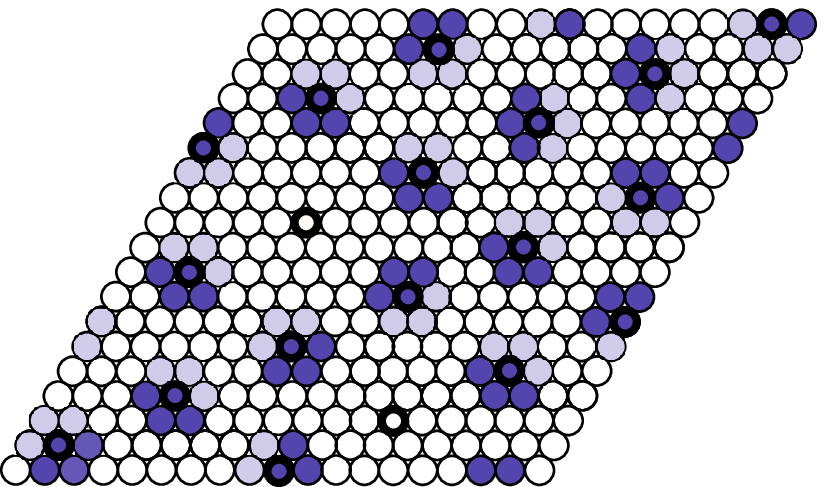}
\includegraphics[scale=1.0]{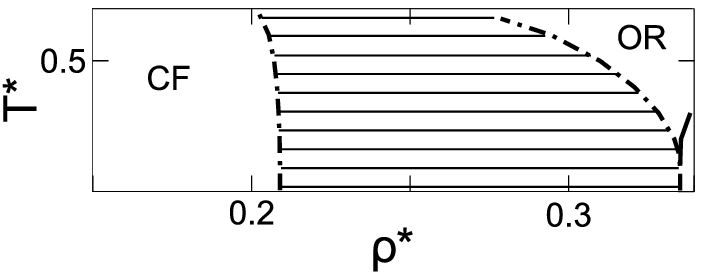}
\caption{A: The black solid circles represent sites of the sublattice considered in sec. 5 A. 
The corresponding sites are occupied if they belong to a rhomboidal cluster. 
The dark blue color indicates one of the possible orientations of the rhomboidal clusters.
B: A portion of the ($\rho^*$,$T^*$) phase diagram for $J^*=3$  with the coexistence region between the  cluster fluid (CF)
and the OR phase (dash-dotted lines) estimated on the basis of the sublattice shown in panel A.
}
\label{sublattice}
\end{figure}

 For $J^*>1$ and the state points that satisfy  $\beta^*(5+4\mu^*)\to 0^-$
all the microscopic states consisting of $N$ noninteracting rhombuses 
are almost equally probable, since the probability of such states
is proportional to $ \exp[\beta^*(5+4\mu^*)N]\approx 1$, and 
for $\beta^*\gg 0$ other states (with $H^*>0$) are rare.
We can obtain an upper bound for the grand potential of the cluster fluid by considering a subset of all such
 microscopic states.  Let us consider the
 sublattice shown in  Fig.\ref{sublattice}a.
The sites of the sublattice can be empty, or occupied by noninteracting rhombuses.
There are 6 possible orientations of the rhombuses at each site of the sublattice and the number of 
sublattice sites is $L^2/19$. The
sublattice  sites are occupied or empty independently of one another, and 
the grand potential and the average density can be obtained immediately,
\begin{equation}
\label{Omsl}
 \Omega^*/L^2=-\frac{T^*}{19}\ln \Bigg(1+6e^{\beta^*(5+4\mu^*)}
\Bigg)
\end{equation}
and
\begin{equation}
 \bar\rho=\frac{4}{19} \frac{6e^{\beta^*(5+4\mu^*)}}{1+e^{\beta^*(5+4\mu^*)}}.
\end{equation}
For $\beta^*(5+4\mu^*)=0$ we obtain $\bar\rho\approx 0.18$. 
This gives the order of magnitude of the density in the cluster fluid for  $\mu^*= -5/4$. 
We compare the grand potential
per lattice site, Eq. (\ref{Omsl}), with 
$\Omega_{MF}^*/L^2$ calculated in the MF approximation for the OR phase. 
This way we obtain a rough 
estimate of the coexistence region between the cluster fluid and the OR phase. 
The corresponding portion of the phase diagram
is shown in Fig.\ref{sublattice}b. Since only a subset of the microscopic states was considered and
  in the disordered phase the 
positions of the rhombuses are not restricted to the  sublattice sites,
the stability region of the CF phase is expected to be larger than shown
 in Fig.\ref{sublattice}b.
By analogy we expect that the molten lamella phase found 
at the coexistence between the OR and lamellar phases for $T^*=0$ will remain stable for $T^*>0$
for the state points that satisfy $\beta^*(2\mu^*+5-6J^*)\to 0$.

\subsection{The effect of mesoscopic fluctuations on the correlation function} 
In this subsection we investigate the effect of mesoscopic fluctuations on the structure
of the disordered phase in the field-theoretic formalism.
 The grand potential (\ref{Omdef}) in 
the coarse-grained description
\cite{ciach:12:0,ciach:11:0,ciach:08:1} is approximated by 
\begin{equation}
\label{Om}
 \beta\Omega \approx - \ln \Big(\int D\rho e^{-\beta \Omega_{MF}[\rho]}\Big)=\beta\Omega_{MF}[\bar\rho]-\ln\Big(
 \int D\phi e^{-\beta H_f[\bar\rho,\phi]}
  \Big)
\end{equation}
where 
\begin{equation}
\label{Hf}
\beta  H_f[\bar\rho,\phi]= \beta\Omega_{MF}[\bar\rho+\phi]-\beta\Omega_{MF}[\bar\rho],
\end{equation}
 $\phi({\bf x})$ is a mesoscopic  fluctuation of the density, and
 the average density  denoted by $\bar\rho({\bf x})$ satisfies the extremum condition
 \begin{equation}
 \label{eqilcon}
  \frac{\delta \beta\Omega}{\delta\rho({\bf x})}=\frac{\delta \beta\Omega_{MF}}{\delta\rho({\bf x})}
  +\langle \frac{\delta \beta H_f}{\delta\rho({\bf x})}\rangle_f=0.
\end{equation}
 In derivation of the above Eq.(\ref{Om}) was used, and $\langle ...\rangle_f$ denotes averaging with the probability 
$\propto \exp(-\beta H_f)$. The MF approximation is valid when the second term on the RHS of (\ref{eqilcon}) is 
negligible. In this theory  $\rho({\bf x})$ is the mesoscopic density, i.e. the density
averaged over a mesoscopic region around each point ${\bf x}$. In Eq.(\ref{Om}) 
$\Omega_{MF}[\rho]$ describes the grand potential 
of  a  system whose mesoscopic density
is constrained to have the form $\rho({\bf x})$ \cite{ciach:12:0,ciach:11:0,ciach:08:1}. In our approximation the 
explicit expression for $\Omega_{MF}[\rho]$ 
is given in Eq.(\ref{wzomeg}). 
We limit our attention to the disordered phase with $\bar\rho({\bf x})=\bar\rho=const.$.
The correlation function in Fourier representation, 
$\langle \tilde \phi({\bf k}) \tilde \phi(-{\bf k})\rangle=\tilde G({\bf k})=1/\tilde C({\bf k})$, 
can be obtained from the equation 
\begin{equation}
\label{Cdef}
 \tilde C({\bf k})=\frac{\delta ^2 \beta\Omega}{\delta\tilde \rho({\bf k})\delta\tilde\rho(-{\bf k})}
= \tilde C_{MF}({\bf k})
 +\langle \frac{\delta ^2 \beta H_f}{\delta\tilde \rho({\bf k})\delta\tilde\rho(-{\bf k})}\rangle_f
 -\langle \frac{\delta \beta H_f}{\delta\tilde \rho({\bf k})}
 \frac{\delta \beta H_f}{\delta\tilde \rho(-{\bf k})}\rangle_f^{conn}
\end{equation}
where
$\langle AB\rangle_f^{conn}=\langle AB\rangle_f-\langle A\rangle_f\langle B\rangle_f$.

 We write (\ref{Hf}) in the form
\begin{equation}
\label{Hfa}
 \beta  H_f[\bar\rho,\phi]= \frac{1}{2}\int d{\bf k} \tilde \phi({\bf k})\tilde C(\bar\rho,{\bf k}) \tilde  \phi(-{\bf k})
 +\Delta H_f,
\end{equation}
 neglect the correction term $\Delta H_f$ and  approximate
the $n-$th functional derivative of $\Omega$ for $n\ge 3$ by the corresponding derivative of
 $\Omega_{MF}$.  This way we obtain the self-consistent Gaussian approximation for the correlation
 function $\tilde G({\bf k})$.
In this self-consistent approximation Eq.(\ref{Cdef}) takes the form
\begin{equation}
\label{C43}
  \tilde C({\bf k})=\beta\tilde V({\bf k})+A_2 +\frac{A_4}{2}\int \frac{d {\bf k}' }{(2\pi)^d \tilde C({\bf k}')} 
-\frac{A_3^2}{2}\int \frac{d {\bf k}' }{(2\pi)^d \tilde C({\bf k}') \tilde C({\bf k}'+{\bf k})}
\end{equation}
where
\begin{equation}
\label{An}
 A_n=\frac{d^n \beta f_h(\rho)}{d\rho^n}
\end{equation}
and for the lattice models $f_h(\rho)$ is given in  Eq.(\ref{fh}).

\begin{figure}
 \includegraphics[scale=1]{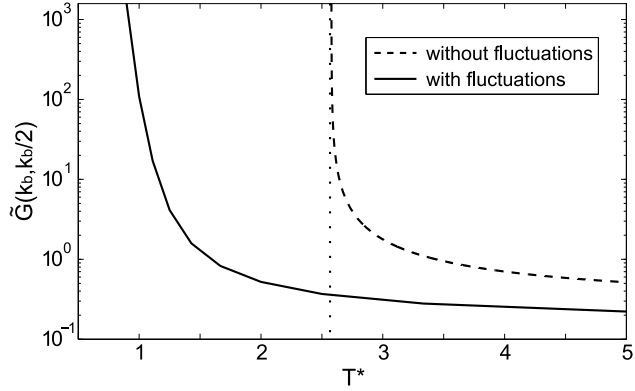}
 \caption{ The maximum of the correlation function in Fourier representation, 
 $\tilde G(k_b,k_b/2)$ for $\rho=1/2$ as a function of the reduced temperature $T^*=k_BT/J_1$. 
 Dashed and solid lines correspond to the MF and  the Brazovskii-type approximation respectively.
 In the MF approximation  $\tilde G(k_b,k_b/2)$ is approximated by $\tilde G_{MF}(k_b,k_b/2)$ 
(see sec. 4 A) which diverges at the $\lambda$-line $T^*= 2.575$, 
while in a presence of fluctuations $\tilde G(k_b,k_b/2)$
  diverges at $T^*=0$.  }
 \label{corr_F_B}
\end{figure}

 Here we limit ourselves to  $\rho\approx 1/2$ where $A_3\approx 0$.
For $A_3=0$ 
  we obtain from (\ref{C43}) the Brazovskii approximation~\cite{brazovskii:75:0},
\begin{equation}
\label{ca}
 \tilde C({\bf k})=\beta\tilde V({\bf k})+A_2 +a(\beta^*,\rho)
\end{equation}
 where 
\begin{equation}
\label{a}
 a(\beta^*,\rho)=\frac{A_4}{2}\int \frac{d {\bf k} }{(2\pi)^d (\beta\tilde V({\bf k})+A_2 +a(\beta^*,\rho))}.
\end{equation}
We  numerically  solve Eq.(\ref{a}) for $\rho=1/2$ and a range of $T^*$.  The maximum of $\tilde G({\bf k})$
is compared with the MF result in Fig.\ref{corr_F_B}.

\begin{figure}
 \includegraphics{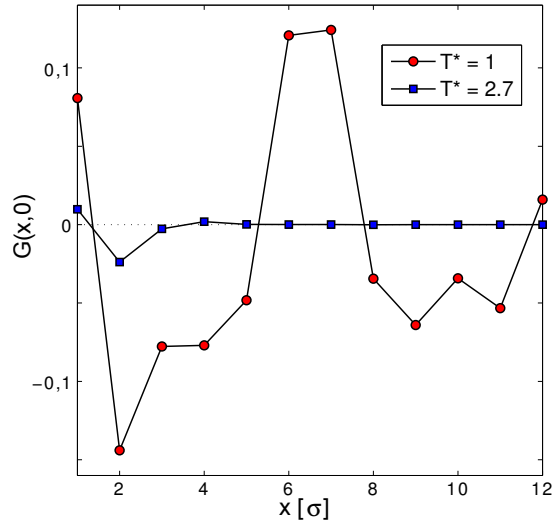}
 \caption{The correlation function
  along the lattice direction, i.e. for the  sites ${\bf x}=x{\bf e}_1$  for  $\rho=1/2$ and two temperatures,
  $T^*=1$ (red circles)   and $T^*=2.7$ (blue squares) in  the Brazovskii approximation. }
 \label{cor_real_B}
\end{figure}

The boundary of stability of the disordered phase in this approximation should be given by $\tilde C({\bf k}_b)=0$.
However, in a  2d system the integral in Eq.(\ref{a}) diverges for the state points for which 
 $\tilde C({\bf k}_b)=0$. Thus, the assumption that
 LHS of Eq. (\ref{ca}) vanishes, leads to divergent RHS
 for $T^*>0$. This indicates the absence of the instability of the disordered phase with respect to 
small-amplitude density waves for $T^*>0$, and a
fluctuation-induced first-order phase transition  is expected~\cite{brazovskii:75:0}. Note, however 
that $\tilde G(k_b,k_b/2)$
is very large, $\sim 10^2$ for $T^*=1$ and very quickly increases for $T^*<1$ (see Fig.\ref{corr_F_B}). 
The correlation function  in the Brazovskii
approximation is shown in Fig.\ref{cor_real_B} for $\rho=1/2$ and two temperatures, $T^*=1$ and $T^*= 2.7$.

\section{Summary and discussion}

We have introduced a triangular lattice model for self-assembly of nanoparticles or
 proteins on surfaces, interfaces or membranes. We have assumed nearest-neighbor attraction and  third-neighbor 
 repulsion. Such interaction ranges were found for example for lysozyme molecules in water \cite{shukla:08:0}.
The advantage of the lattice model is the possibility of detailed investigation of the ground state,
 where the disordering effect of thermal motion is absent. We have found stability regions
 of periodically distributed clusters, bubbles or stripes. A very interesting property of the ground state is its 
strong degeneracy at the coexistence between ordered phases 
of different symmetry. The entropy per site at the coexistence between different ordered phases
does not vanish. 
We identify the disordered states stable at the phase coexistence between vacuum  and hexagonally ordered clusters 
(Fig.\ref{gsc}a,c)
with a disordered  cluster fluid.
 The disordered states stable at the coexistence between the ordered 
clusters and stripes (Fig.\ref{gsc}e-j) correspond to a disordered phase called molten lamella. 
 The structures stable at
 the above phase coexistences include the interface between the two phases. Thus, the surface tension vanishes.

The vanishing surface tension and the strong degeneracy of the GS at the phase coexistence as well as an
ultra-low surface tension
for $T^*>0$ were observed previously
  in surfactant solutions \cite{ciach:89:0,ciach:01:2,gompper:94:3}. 
 Here we show that the amphiphilic molecules are not necessary to obtain the vanishing
  surface tension at $T=0$. The surface tension vanishes when  the ground state is strongly degenerated 
at the phase coexistence, and the stable structures include the interface.
The effects of the vanishing surface tension in the case of the colloid and amphiphilic self-assembly are very similar.
Namely, disordered phases with strong local inhomogeneities become 
stable.  In the case of amphiphiles these phases are the micellar, the microelmulsion and the sponge  phases, 
whereas in the case of the SALR potential - the cluster fluid and the molten lamella. 

The degeneracy of the ground state leads to a huge number of thermodynamically stable patterns.
Particularly interesting are the patterns stable in the molten lamella phase. The patterns are composed of a 
few motifs: clusters, stripes or rings that are surrounded by a single layer of empty sites.
A transition between different stable patterns is a collective phenomenon, involving a large fraction of 
the particles. The information encoded in the stable patterns cannot be easily destroyed by the 
thermal motion when  $T^*$ is not high.

 We have determined the phase diagram for $T^*>0$ in the mean-field approximation. 
We have obtained the same sequence of the orderd periodic phases as in Ref.\cite{archer:08:0}. For increasing density 
the stable phases are: fluid, hexagonally
ordered clusters, lamellar phase and hexagonally ordered bubbles. For high $T^*$ the phase
 diagrams on the lattice and in continuum are very similar. However, on the lattice 
 there are two lamellar phases with different 
orientations of the stripes w.r.t the lattice directions. For strong repulsion ($J^*=3$) we obtain 
two hexagonal phases, with and without orientational ordering of the long axes of the rhomboidal clusters. 
The regions occupied by the ordered phases and the extent of the two-phase regions
 on the $(\rho^*,T^*)$ phase diagram are also different than in Ref.\cite{archer:08:0}.
For weak repulsion 
($J^*=1$, i.e. $\int d{\bf r}V(r)=0$) the clusters are symmetrical and there is a single haxagonal phase, as 
in Ref.\cite{archer:08:0}.

Unfortunately, in the mean-field approximation the effect of the degeneracy of the ground state cannot be correctly 
described. In mean field the mesoscopic fluctuations are neglected, whereas in the SALR systems
 the  displacements of the clusters or stripes 
lead to formation of the disordered cluster fluid or molten lamella. 
 The dominant role of mesoscopic fluctuations makes the studies
of the SALR systems particularly difficult. In off-lattice systems the
phase diagrams obtained in the DFT \cite{archer:08:0} and in MC simulations \cite{imperio:06:0} 
differ significantly from each other. The main features present on the MC and absent on the DFT phase diagram
 are: (i) a stability of the cluster fluid phase 
 between the ordered cluster
 phase and the homogeneous fluid (ii)  a reentrant melting of the ordered cluster phase for high temperatures.
Similar difference between the high-$T$ part of the phase diagrams in mean-field  approximation and in 
a presence of fluctuations was observed   in the context of block copolymers  when fluctuations were taken 
into account 
within field-theoretic methods \cite{brazovskii:75:0,fredrickson:87:0,podneks:96:0}.

 We expect similar differences between mean-field and exact results for the present lattice model.
From our preliminary studies of the effects of fluctuations  two conclusions follow:
(i) the cluster fluid  is stable for  low $T^*$ up to the density $\rho=0.2$ or larger, and (ii)
  the transition to the lamellar phase is fluctuation-induced first order. The fluctuation-induced first-order 
transitions are usually very weakly first order. Moreover, the maximum of the structure factor
increases to very large values for $T^*<0.8$, and it may be difficult to determine the order of the 
transition in experiment or simulation.
The effects of fluctuations on the phase diagram of this model 
are studied in much more detail by Monte Carlo simulations in Ref.\cite{almarza:14:0}.

{\bf Acknowledgments}
The work of JP  was   realized within the International PhD Projects
Programme of the Foundation for Polish Science, cofinanced from
European Regional Development Fund within
Innovative Economy Operational Programme "Grants for innovation".
 AC and JP acknowledge the financial support by the NCN grant 2012/05/B/ST3/03302. 
 N.G.A. gratefully acknowledges financial support from the Dirección General de Investigación Científica y Técnica under Grant No.
FIS2010-15502, from the Dirección General de Universi-
dades e Investigación de la Comunidad de Madrid under
Grant No. S2009/ESP-1691 and Program MODELICO-CM.

\section{Appendix. Grand potential for weakly ordered periodic phases in MF}
The normalized functions $g_p$  satisfy the equations
\begin{equation}
\label{g1}
 \frac{1}{V_u}\sum_{{\bf x}\in V_u}g_p({\bf x})=0
\end{equation}
and
\begin{equation}
\label{g2}
  \frac{1}{V_u}\sum_{{\bf x}\in V_u}g_p^2({\bf x})=1.
\end{equation}
In the above equations the summation is over the unit cell of the ordered structure with the area $V_u$. 
The length of the unit cell in our case is $2\pi/k_b$. 
The functions $g_p$ for the lamellar and  hexagonal phases have the forms 
\begin{equation}
 g_l({\bf x})=\sqrt{2} \cos(k_b {\bf x}\cdot{\bf e}_i)
\end{equation}
and
\begin{equation}
 g_h({\bf x})=\sqrt{\frac{2}{3}}\sum_{i=1}^3 \cos(k_b {\bf x}\cdot{\bf e}_i)
\end{equation}
where $k_b$ is given in Eq.(\ref{kb}). In the case of noninteger $2\pi/k_b$, in order to calculate Eqs.(\ref{g1})
 and (\ref{g2}), we make the approximation 
\begin{equation}
 \frac{1}{V_u}\sum_{{\bf x}\in V_u} f(k_b{\bf x})\simeq 
 \int_0^{2\pi}\frac{dz_1}{(2\pi)}\int_0^{2\pi} \frac{dz_2}{(2\pi)}f({\bf z}),
\end{equation}
 where $z_i=k_bx_i$.

From the condition $\partial \Delta\Omega/\partial \delta\rho_p=0$ we obtain for $\Phi_p \ll 1$
\begin{equation}
 \delta \rho_p \cong - \frac{A_3 \Phi_p^2}{2(\beta^{*} {\tilde{V}^{*}(0) + A_2)}},
\end{equation}
where $A_n$ is defined in Eq.(\ref{An}),
and $\beta^{*}\tilde{V}^{*}(0) \rho_0 + A_1 - \beta^{*} \mu^{*} = 0$. After some algebra we obtain 
the approximate expression
\begin{equation}
\beta^{*} \Delta \Omega_p^{*} = \Phi^2_p \,\, \frac{\beta^{*} \tilde{V}^{*}(\mathbf{k}_b)+A_2}{2} +
 \Phi^3_p \,\, \frac{A_3 \kappa_3^{p}}{3!}
+ \Phi^4_p \Big[\frac{A_4\kappa_4^p}{4!}-\frac{A_3^2}{8(\beta^{*} \tilde{V}^{*}(0)+A_2)}\Big]+O(\Phi^5_p),
\label{gwiazdka}
\end{equation}
where the geometric factors  are defined as $\kappa_n^{p}=\frac{1}{V_u}\sum_{V_u}g_p({\bf x})^n$
 and take the following
values $\kappa_3^l = 0$, $\kappa_4^l = \frac{3}{2}$, $\kappa_3^h = \sqrt{\frac{2}{3}}$, $\kappa_4^h = \frac{5}{2}$
\cite{ciach:10:1}.
From  $\partial \Delta\Omega/\partial \delta\Phi_p=0$ we obtain the amplitude $\Phi_p$, and after inserting it
 to (\ref{gwiazdka}), the 
value of $\beta \Delta \Omega_p^*$ for given $T^*$ and $\rho_0$.


\end{document}